\documentclass{elsart}

\usepackage{natbib}

 \usepackage{graphics}

\usepackage{amssymb}

\begin{document}

\begin{frontmatter}



\title{ \bf Recent NA48 results on semileptonic and rare kaon decays. Measurement of $V_{us}$. }


\author{ L. Litov\thanksref{label1}}
\address{CERN, Geneva, Switzerland}
\ead{ e-mail: Leandar.Litov@cern.ch}
\thanks[label1]{For the NA48 Collaboration}
\author{Talk given at ICHEP04, Beijing, 2004}

\begin{abstract}
$\qquad$The new results of investigation of $K^0_Le3$ and $K^{\pm}e3$  decays performed by NA48 Collaboration are presented. The measured branching fractions are used to extract $V_{us}$. The values of Ke3 form-factors and radiative $K^0_Le3\gamma$ branching ratio are reported. A  list
of recent NA48 results concerning some rare neutral kaon decays is given.   
\vspace{1pc}

\end{abstract}




\end{frontmatter}




\section{Introduction}
The unitarity condition for the quark-mixing CKM matrix leads to a number of relations between its elements. Thus for the first row one has $|V_{ud}|^2+|V_{us}|^2+|V_{ub}|^2 = 1$. The contribution of $|V_{ub}|^2 \simeq 10^{-5}$ is negligible. The value
 of $|V_{ud}|= 0.9738 \pm 0.0005$  is well determined from 
superallowed nuclear decays and measured free neutron life time\cite{PDG}. 
From the  unitarity relation  we obtain  $|V_{us}|= 0.2274 \pm 0.0021$ in 2.2 
$\sigma$ disagreement  with the experimental value\cite{PDG}  $|V_{us}|= 0.2200 \pm 0.0026$. 
The most precise information on $|V_{us}|$ comes from measurement of the branching fraction of $Ke3$ decays. The NA48 Collaboration has performed a high precision investigation of  $K^0_Le3$ and $K^{\pm}e3$ decays. Based on it we extract the value of the $V_{us}$ element. The recent NA48 results of measurement of branching fractions and form-factors of some rare $K^0_L$ and $K^0_s$ decays are reported as well.  
   
\section{Experimental setup}
The NA48 setup\cite{NA48S} is designed to measure direct CP violation and rare decays in the Kaon system. The beam line is  able to produce simultaneous $K^0_L$ and $K^0_s$ beams with momentum (20 $\div$  200) GeV/c. The detector consists of 90 m long evacuated decay tube, followed by the charged particle spectrometer including four drift chambers fitted in helium tank, each with 8 planes of sense wires. The magnet with a field integral of 0.85 Tm is placed after the first two chambers. The scintillator hodoscope with time resolution of 200 ps is situated after the helium tank in front of quasi-homogeneous electromagnetic calorimeter (LKr) based on liquid Krypton. An iron-scintillator hadron calorimeter complements the krypton calorimeter in measuring the energy of hadrons. The muon veto system consisting of scintillator counters, shielded by iron walls is used to identify muons. In 2003 the beam line was upgraded to transport simultaneously positive and negative particles with central momentum of 60 GeV/c to the NA48/2 detector\cite{kekel}. 
\section{Investigation of semileptonic Ke3 decays}
\subsection{$K^0_L \rightarrow \pi e \nu$ branching ratio}\label{subsec:prod}
 The data were taken during a special minimum bias run with $K^0_L$ beam in 1999. The trigger requirements were two charged particles in the scintillator hodoscope or in the drift chambers coming from a vertex in the decay region.  We measure the ratio of the widths of $K^0_Le3$  decay and all $K^0_L$ decays with two charged particles in the final state\cite{Ke3}. The branching ratio of the two-track decays is experimentally known and is given by
\begin{equation} 
Br(2tr)=1.0048 - Br(K^0_L \rightarrow 3\pi^0)
\label{eq:b2t}. 
\end{equation}
In the basic selection, two in-time tracks ($\delta t < 6 ns$) with opposite charges and momenta $> 10 GeV/c$, defining a vertex inside the fiducial decay volume were required. An additional condition on the sum of the two track momenta to be higher then 60 GeV/c was applied. In this way 12.6 million two-track events were selected. The $Ke3$ signal was separated from two-track sample by requiring that at least one of the tracks is consistent with an electron. Thus 6.76 million events were accepted. This number was corrected for the inefficiency of the electron identification and for the background coming from $K\mu3$ and $K2\pi$ decays due to misidentification of pions as electrons. \\
The detector response was simulated for the five two-track decay modes  $\pi e \nu$, $\pi \mu \nu$, $\pi^+ \pi^-$, $\pi^0 \pi^+ \pi^-$ and $\pi^0 \pi^0 \pi^0$ with Dalitz decay of one $\pi^0$. The radiative corrections with real and virtual photons were included in the Ke3 decay simulation\cite{Ginsb}. The average two-track acceptance was calculated as a  weighted mean of the individual acceptances using the corresponding  branching ratios obtained as a weighted average of the PDG values and the recent KTeV measurement\cite{KTeVb}.
The error in 
\begin{equation}             
\frac{Br(K^0_Le3)}{Br(2tr)} = 0.4978 \pm 0.0035  
\label{eq:R}
\end{equation}
is dominated by the uncertainties of the $K_L$ momentum spectrum and two-track acceptance. Statistic error is negligible.
The value  $Br(K^0_L3\pi^0) = (19.92 \pm 0.70)\%$ used to obtain
\begin{equation}             
Br(K^0_Le3)= 0.4010 \pm 0.0028_{exp} \pm 0.0035_{norm}  
\label{eq:ke3}
\end{equation}
was determined as the weighted mean of PDG value $(21.05 \pm 0.28)\%$ and the recent KTeV result\cite{KTeVb} $(19.45 \pm 0.18)\%$. Our value of $Br(K^0_Le3)$ exceeds the PDG one by 2.5$\sigma$  and is in agreement with recent KTeV\cite{KTeVb} and KLOE\cite{KLOE} measurements.
\subsection{$K^0_L \rightarrow \pi^0 \pi^0 \pi^0$ branching ratio}\label{subsec:prod}
The dominant uncertainty in our result for Br(Ke3) comes from the Br($K^0_L3\pi^0)$. The two existing measurements\cite{Kreitz,KTeVb} of this decay are inconsistent by $\sim\!5\sigma$. In order to clarify the situation we measured the $K^0_L3\pi^0$ branching fraction using NA48/1 data obtained in high-intensity $K^0_s$ beam run in 2000. The magnet spectrometer drift chambers were removed from the experimental set-up thus leading to absence of any material between the last beam collimator and the electromagnetic calorimeter. The  $K^0_L3\pi^0$ branching fraction was extracted from the measurement of the ratio $Br(K^0_L \rightarrow \pi^0 \pi^0 \pi^0)/Br(K^0_s \rightarrow \pi^0 \pi^0)$ advancing from the good knowledge\cite{PDG} of $Br(K^0_s \rightarrow \pi^0 \pi^0) = 0.3104 \pm 0.0014$. About 200 000 $K^0_L \rightarrow \pi^0 \pi^0 \pi^0$ and 600 000 $K^0_s \rightarrow \pi^0 \pi^0$ were fully reconstructed using only a small fraction of available data. The main uncertainties in this measurement are induced by the LKr energy scale, effective target position and $K^0_L$ life time. Our preliminary result      
\begin{equation}             
Br(K^0_L3\pi^0) = 0.1966 \pm 0.0006_{stat} \pm 0.0033_{syst}  
\label{eq:3pi0}
\end{equation}
is in a good agreement with KTeV  one and with the preliminary KLOE measurement\cite{KLOE}. Implementation of (\ref{eq:3pi0}) in (\ref{eq:b2t}) leads to a slight change of $0.3\%$ in our value for Br($K^0_Le3$). 
\subsection{$K^\pm \rightarrow \pi^0 e^\pm \nu$ branching ratio}\label{subsec:prod}
For this measurement we analyzed  data collected in 2003 by NA48/2 setup in a special low-intensity run in simultaneous 60 GeV kaon beams with opposite charges. A minimum bias trigger requiring at least one charged particle hitting the scintillator hodoscope was employed. As a normalization channel we used the decay $K^\pm \rightarrow \pi^\pm \pi^0$. Its branching ratio  $Br(K^\pm \rightarrow \pi^\pm \pi^0) = 0.2113 \pm 0.0014$ is  measured precisely enough in several experiments. Some 59 000 $K^+ \rightarrow \pi^0 e^+ \nu$, 33 000 $K^- \rightarrow \pi^0 e^- \nu$, 468 000 $K^+ \rightarrow \pi^+ \pi^0$ and 260 000 $K^- \rightarrow \pi^- \pi^0$ events were selected with negligible  background. The radiative corrections with virtual and real $\gamma$'s were taken into account in the simulation of Ke3 decay. The main sources of systematic uncertainties were identified to be the detector acceptance, the normalization and MC statistics. The following preliminary values for Ke3 branching ratios were found: $$\begin{array}{l}
Br(K^+e3) = (5.163 \pm 0.021_{stat} \pm 0.056_{syst})\%\\[4pt]
Br(K^-e3) = (5.093 \pm 0.028_{stat} \pm 0.056_{syst})\%\\[4pt]
Br(K^\pm e3) = (5.14 \pm 0.02_{stat} \pm 0.06_{syst})\%
\end{array}$$
Our results confirm  E865 measurement\cite{BNL} and considerably disagree with the value suggested by PDG. 
\subsection{Extraction of $|V_{us}|$}\label{subsec:prod}
The $|V_{us}|$ element can be extracted from the Ke3 decay parameters\cite{Cir}
\begin{equation}
|V_{us}|f^{K\pi}_+(0) = \sqrt{\frac{128\pi^3\Gamma(Ke3(\gamma))}{C^2G^2_FM^5_KS_{EW}I_K(\lambda_+)}}
\end{equation}
where $S_{EW}= 1.0232$ is the short distance enhancement factor, $C=1$ for $K^0$ and $C=1 / \sqrt{2}$ for $K^\pm$, $I_K(\lambda_+)$ is the phase space integral and $f^{K\pi}_+(0)$ is the $Ke3$ form factor at zero transfer momentum. To evaluate $|V_{us}|$ we followed the prescription and used the numerical results of Ref.10. A linear parameterization of $f_+(t)$ with corresponding PDG $\lambda_+$ values was employed for calculation of  $I_K(\lambda_+)$. The integration was performed over the Ke3 Dalitz plot, corrected for the loss of radiative $Ke3\gamma$ events. Using Eq.(5) and PDG values for the $K^\pm$ and $K^0_L$ mean life times we find $$
\begin{array}{l}
|V_{us}|f^{K^0\pi^+}_+(0) = 0.2146 \pm 0.0016\\[4pt]
|V_{us}|f^{K^+\pi^0}_+(0) = 0.2250 \pm 0.0013\\[4pt]
|V_{us}|f^{K^-\pi^0}_+(0) = 0.2235 \pm 0.0014\\[4pt]    
|V_{us}|f^{K^\pm\pi^0}_+(0) = 0.2245 \pm 0.0013
\end{array}$$
These results, together with the recent experimental measurements, PDG values and SM predictions obtained using $f^{K^0\pi^+}_+(0)= 0.981 \pm 0.010$ and  $f^{K^+\pi^0}_+(0)= 1.002 \pm 0.010$ are presented on Figure 1 . The form factors were  calculated\cite{Cir} in the framework of $\chi PT$ taking into account the isospin violation effects, electromagnetic corrections and $O(p^6)$ terms. For $|V_{us}|$ we obtain      
\begin{figure}[hbtp]
  \begin{center}
   \rotatebox{-90}{
    \resizebox{6cm}{!}{\includegraphics{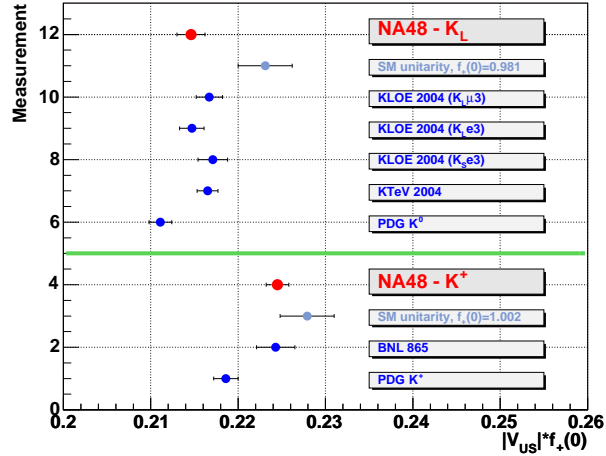}}}
    \caption{$|V_{us}|f^{K\pi}_+(0)$. The SM predictions were obtained using the corresponding values for $f^{K\pi}_+(0)$ from Ref.10 }
    \label{fig:fig1}   
  \end{center}
\end{figure}
$$
\begin{array}{l}
|V_{us}|^{K^0\pi^+} = 0.2187 \pm 0.0016_{exp} \pm 0.0023_{th}\\[4pt]
|V_{us}|^{K^+\pi^0} = 0.2246 \pm 0.0013_{exp} \pm 0.0023_{th}\\[4pt]
|V_{us}|^{K^-\pi^0} = 0.2231 \pm 0.0014_{exp} \pm 0.0023_{th} \\[4pt]    
|V_{us}|^{K^\pm\pi^0} = 0.2241 \pm 0.0013_{exp} \pm 0.0023_{th}
\end{array} $$
 The errors on $|V_{us}|$ are dominated by the theoretical uncertainty in the calculation of $O(p^6)$ contribution to $f^{K^+\pi^0}_+(0)$. The values of  $|V_{us}|$ extracted from measurement of charged $Ke3$ decays are in disagreement with PDG value and in good agreement with E865 result and SM prediction confirming the CKM unitarity. Similarly in the $K^0$ sector, our result is significantly higher then PDG value but in good agreement with recent KTeV and KLOE measurements. However the value of $|V_{us}|$ is still 2.5$\sigma$ away from SM prediction.
\subsection{Other results on Ke3 decays}\label{subsec:prod}
A sample of 5.6 million reconstructed events from the same 1999 run was used to measure the Dalitz plot density of $K^0_Le3$ decay\cite{na48ff}. Admitting all possible Lorentz-covariant couplings, the form factors for vector ($f_+(q^2)$), scalar ($f_s$) and tensor ($f_T$) interactions were measured. The linear slope of the vector form factor $\lambda_+ = 0.0284 \pm 0.0007 \pm 0.0013$ and the values of the ratios $|f_s/f_+(0)| = 0.015^{+0.007}_{-0.010} \pm 0.012$ and  $|f_T/f_+(0)| = 0.05^{+0.03}_{-0.04} \pm 0.03$ were obtained. The values of $f_s$ and $f_T$ are consistent with zero. Assuming only vector couplings, $\lambda_+ = 0.0288 \pm 0.0005 \pm 0.0011$ was measured. No evidence for a quadratic term $\lambda^{''}$ different from zero was found. Alternatively, a fit to a dipole form factor yields a pole mass $M = 859 \pm 18 MeV$, consistent with the $K^*(892)$ mass.\\
The relative branching ratio of the decay $K^0 \rightarrow \pi^\pm e^\mp \nu\gamma$ with respect to  $K^0 \rightarrow \pi^\pm e^\mp \nu$ decay was measured\cite{na48rad}. The result $Br(K^0e3\gamma,E^*_\gamma>30MeV,\theta^*_{e\gamma}>20^o)/Br(K^0e3) = (0.964 \pm 0.008^{+0.012}_{-0.011})\%$ is based on 19 000 $Ke3\gamma$ and $5.6.10^6$ Ke3 decays. It agrees with recent theoretical predictions\cite{Gasser} but is at variance with the KTeV result\cite{KTeVr}.
\section{Rare $K^0$ decays}
 Based on the investigation of 60 000 $K^0_L \rightarrow e^+e^-\gamma$ decays, we measured the BMS form factor $\alpha_K^* = -0.207 \pm 0.019 \pm 0.017$ in good agreement with KTeV result.\\
About 200 events of the rare double Dalitz decay $K_L \rightarrow e^+e^-e^+e^-$ were observed leading to the preliminary value for $Br(K_L \rightarrow e^+e^-e^+e^-) = (3.30 \pm 0.24_{stat} \pm 0.14_{syst} \pm 0.24_{norm}).10^{-8}$.\\
The decay $K_L \rightarrow \pi^\pm \pi^0 e^\mp \nu$ was measured with an improved precision\cite{ke4}: $Br(K_L \rightarrow \pi^\pm \pi^0 e^\mp \nu) = (5.21 \pm 0.07_{stat} \pm 0.09_{syst}).10^{-5}$. The form factors were determined by fitting the differential cross sections to the  Cabbibo-Maksymowich variables. The result agrees with previous measurements and with theoretical predictions.\\
A search for the CP violating decay $K_s \rightarrow 3\pi^0$ was performed\cite{3pi0}. Fitting the life time distribution of about 4.9 million reconstructed $K^0/\overline{K^0}\rightarrow3\pi^0$ decays, the CP violating amplitude was found to be\\ $Re(\eta^{000}) = -0.002 \pm 0.011_{stat} \pm 0.015_{syst}$; \\$Im(\eta^{000}) = -0.003 \pm 0.013_{stat} \pm 0.017_{syst}$\\ leading to an upper limit on the $Br(K_s \rightarrow 3\pi^0) < 7.4.10^{-7}$ at $90\%$ confidence level.\\
Exploiting the data collected by NA48/1 collaboration during 2002 with a high-intensity $K_s$ beam,  6 $K_s \rightarrow \pi^0\mu^+\mu^-$ decays were observed\cite{pimumu} with a background expectation of $0.22^{+0.18}_{-0.11}$ events. The measured branching ratio is $Br(K_s \rightarrow \pi^0\mu^+\mu^-)=(2.9^{+1.5}_{-1.2 stat} \pm 0.2_{syst}).10^{-9}$.

\end{document}